\documentclass[compsoc,conference,a4paper,10pt,times]{IEEEtran} %EUROSP

\IEEEoverridecommandlockouts
\usepackage{cite}
\usepackage{amsmath,amssymb,amsfonts}
\usepackage{balance}
\usepackage{subfigure}
\usepackage{algorithmic}
\usepackage{graphicx}
\usepackage{textcomp}
\usepackage{xcolor}
\usepackage{bm}
\def\BibTeX{{\rm B\kern-.05em{\sc i\kern-.025em b}\kern-.08em
    T\kern-.1667em\lower.7ex\hbox{E}\kern-.125emX}}

\IEEEoverridecommandlockouts\IEEEpubid{\makebox[\columnwidth]{ 978-1-6654-3540-6/22~\copyright~2022 IEEE \hfill} \hspace{\columnsep}\makebox[\columnwidth]{ }}

\begin{document}

\title{Intelligent Detection of Non-Essential IoT Traffic on the Home Gateway}

\author{\IEEEauthorblockN{Fabio Palmese}
\IEEEauthorblockA{\textit{DEIB, Politecnico di Milano} \\
\textit{Milan, Italy}\\
%fabio.palmese@polimi.it
}
\and
\IEEEauthorblockN{Anna Maria Mandalari}
\IEEEauthorblockA{\textit{University College London} \\
\textit{London, UK}\\
%a.mandalari@ucl.ac.uk
}
\and
\IEEEauthorblockN{Hamed Haddadi}
\IEEEauthorblockA{\textit{Imperial College London} \\
\textit{London, UK}\\
%h.haddadi@imperial.ac.uk
}
\and
\IEEEauthorblockN{Alessandro E. C. Redondi}
\IEEEauthorblockA{\textit{DEIB, Politecnico di Milano} \\
\textit{Milan, Italy}\\
%alessandroenrico.redondi@polimi.it
}
}

\maketitle

\begin{abstract}
The rapid expansion of Internet of Things (IoT) devices, particularly in smart home environments, has introduced considerable security and privacy concerns due to their persistent connectivity and interaction with cloud services. Despite advancements in IoT security, effective privacy measures remain uncovered, with existing solutions often relying on cloud-based threat detection that exposes sensitive data or outdated allow-lists that inadequately restrict non-essential network traffic. This work presents ML-IoTrim, a system for detecting and mitigating non-essential IoT traffic (i.e., not influencing the device operations) by analyzing network behavior at the edge, leveraging Machine Learning to classify network destinations. Our approach includes building a labeled dataset based on IoT device behavior and employing a feature-extraction pipeline to enable a binary classification of essential vs. non-essential network destinations. We test our framework in a consumer smart home setup with IoT devices from five categories, demonstrating that the model can accurately identify and block non-essential traffic, including previously unseen destinations, without relying on traditional allow-lists. We implement our solution on a home access point, showing the framework has strong potential for scalable deployment, supporting near-real-time traffic classification in large-scale IoT environments with hundreds of devices. This research advances privacy-aware traffic control in smart homes, paving the way for future developments in IoT device privacy.
\end{abstract}


\section{Introduction}
\label{sec:intro}
The number of Internet of Things (IoT) devices is increasing dramatically, revolutionizing our daily lives, especially in the smart home field \cite{sepasgozar2020systematic}. However, the presence of these devices introduces several security and privacy challenges.
In recent years, IoT devices have become frequent targets or sources of security threats affecting users and other Internet entities \cite{hassija2019survey}. Given their constant connectivity and communication with cloud services, IoT devices pose significant privacy risks, as users are often unaware of what information is being shared and who is collecting it.
Although many solutions in the literature effectively counter known security threats, such as through device protection and isolation (e.g., DDoS detection), privacy concerns remain inadequately addressed \cite{mandalari2023protected}. Moreover, most existing solutions are cloud-based, leading to further exposure of sensitive data during threat detection processes \cite{mandalari2023protected, safronov2024sunblock}.
Privacy-aware approaches, typically based on allow-lists, restrict network traffic to only pre-approved destinations. However, these methods are often ineffective, as they depend on predefined lists that are frequently incomplete or outdated, particularly when addressing advertisements or user tracking \cite{he2024can}. The analysis of the network behavior of consumer smart devices highlights frequent communication with non-essential destinations, not contributing to device operations.
This work presents ML-IoTrim, a system designed for the home gateway for detecting non-essential IoT traffic by inspecting network characteristics. The goal is to obtain a centralized smart home entity capable of automatically detecting and blocking non-essential traffic to ensure a privacy-aware environment.
Leveraging recent advances in Machine Learning (ML) for network traffic analysis, we propose a binary classification for network destinations. To achieve this, we first outline a methodology to build a labeled dataset by observing IoT device behavior over time and identifying non-essential destinations. Then, we describe a feature-extraction pipeline that generates statistical information from network data that can be used to train supervised  ML models. We present experimental results on a consumer smart home environment with IoT devices deployed in a large testbed. 
We show that a global model can effectively distinguish essential traffic by only relying on network traffic patterns, with no information on domain names or IP addresses. Additionally, we successfully classify new destinations not present in the training set, addressing one of the main limitations of allow-list approaches, requiring frequent updates.
We implement the framework and test it on a mini PC acting as a home access point, showcasing its potential in smart homes. Employing the proposed solution in large-scale environments is possible, classifying traffic from hundreds of devices in real-time.\\
We release ML-IoTrim publicly for reproducible research\footnote{https://github.com/SafeNetIoT/ML-IoTrim}.
%We strongly believe that the proposed solution can enhance user privacy in smart home environments by limiting information exposure and restricting communication only to destinations that are essential for device operations.\\
The remainder of this paper is organized as follows: Section \ref{sec:related} discusses the literature on privacy/security for the smart home, Section \ref{sec:overview} introduces the proposed framework, including the methodology used to obtain the ground truth labels and the setup of the testbed, Section \ref{sec:ml-iotrim} presents the learning module with the data processing pipeline and the ML models involved for destination classification. Experimental results are reported in Section \ref{sec:results}, while in Section \ref{sec:implementation} we discuss the framework integration in a real-life home environment. Section \ref{sec:concl} concludes the work and discusses future research directions.

\section{Related Work}
\label{sec:related}
Several works in the literature tackled the problem of user privacy and security in IoT by proposing tools and frameworks. We comment here on the main work related to privacy for IoT devices, focusing specifically on consumer devices for the smart home. The authors of \cite{habibi2017heimdall} propose an allow-list-based tool to mitigate IoT denial of service attacks in the home gateway. However, considering the continuous updates that such lists require, allow-list-based solutions have been proven ineffective in the dynamic field of the IoT \cite{he2024can}. The work in \cite{lastdrager2020protecting} proposes SPIN, a distributed framework to be deployed in a wide range of home networks, providing a measurement-based data model of IoT devices and their security characteristics. The framework allows visualization and control of IoT traffic, with a privacy manager allowing consumers to control their privacy in insecure devices. A reverse firewall is also implemented to automatically block devices when required, for example, when a DDoS attack is detected. In \cite{bugeja2021prash}, the authors present a framework for early identification of privacy threats in the smart home. The authors present a system model, a threat model, and a set of privacy metrics to describe the status of devices in the network and notify the user of potential issues in the network. Similarly, the authors of \cite{sturgess2018capability} reduce the smart home devices to their data-collection capabilities to state the privacy risks derived by the presence of the devices, depending on the information the user exposes. However, the work only describes the model details, and no implementation is presented. The number of solutions for ensuring IoT security and privacy in the smart home keeps increasing; however, the proposed frameworks usually lack intelligence or automation, leading to inefficient control of IoT security. Besides academic works, different commercial solutions have been released in the last few years to overcome IoT-related issues in smart environments. Different solutions have been released to protect IoT devices from external attackers or to prevent such devices from being the sources of attacks \cite{bitdefender,shieldiot,fing}. Different anti-virus producers also extended their operations with IoT-specific tasks for the smart home \cite{kaspersky}.
Several companies integrated software solutions in custom hardware components to obtain a full "security-by-design" product, usually serving as a smart home hub or gateway \cite{firewalla,rattrap}. Overall, many different solutions promise to secure IoT devices in the smart home. However, all these solutions focus mainly on security attacks, ignoring privacy-related threats, and often do not properly provide what they promise \cite{mandalari2023protected}. The work in \cite{mandalari2021blocking} presented a methodology to retrieve non-essential destinations from an IoT device. However, the pipeline requires manual controls, making it impossible for large-scale deployment. Unlike existing solutions, we provide a framework that can automatically detect non-essential traffic without prior knowledge on the device or the network domains.

\section{Overview}
\label{sec:overview}
\begin{figure}[t]
    \begin{center}
      \includegraphics[width=0.9\columnwidth]{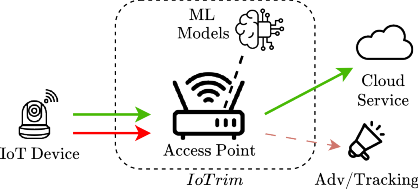}
      \caption{Sketch of the ML-IoTrim framework architecture}
      \label{fig:iotrim_framework}
    \end{center}
\end{figure}
To meet the outlined objectives, we propose ML-IoTrim, a framework implemented on the home gateway to automatically distinguish and block non-required traffic. The framework identifies non-essential destinations by relying solely on network traffic characteristics and blocks them accordingly. Classification is performed using pre-trained machine learning models, which are preloaded onto the access point.
Figure \ref{fig:iotrim_framework} sketches the proposed architecture with the required components. The system continuously monitors network traffic from connected devices and performs periodic analyses to extract features. Machine learning classifiers then make decisions on whether a contacted destination is essential or not for the device so that specific procedures can block traffic toward non-essential destinations. 
First, we present the approach for collecting information on essential and non-essential destinations, forming the ground truth for the network dataset. Next, we describe our testbed setup, followed by the data collection process, while data analysis and machine learning classification will be covered in the next sections.
\subsection{Methodology}
The first step towards developing the framework is to obtain a network dataset with associated information on the destinations categorized as essential or non-essential for the device, to be used to train the machine learning models.
While collecting network traffic can be achieved using tools such as \textit{tcpdump} or \textit{Wireshark}, distinguishing between essential and non-essential destinations poses a greater challenge. IoT device manufacturers typically do not provide documentation on the standard network behavior of their devices, despite the release of recent IETF standards trying to enforce this transparency (e.g., the Manufacturer Usage Description \cite{lear2019rfc}).
To address this, we adapted the pipeline introduced in \cite{mandalari2021blocking}, which analyzes network traffic from consumer IoT devices to distinguish required network destinations. The procedure relies on a custom access point with the MonIoTr framework, able to monitor IoT devices to collect and control their network traffic \cite{ren2019information}.
Each device is analyzed separately using a two-step procedure to (i) retrieve the list of destinations and (ii) distinguish the nature of each destination.
For the first step, we distinguish destinations contacted as the device is booted (Power-on destinations) from task-specific ones contacted during the execution of device operations (Activity destinations).
To collect a complete set of destinations, we repeat the following procedure 10 times:(i) the device is unplugged, (ii) the traffic capture is started, and (iii) the device is plugged and powered on to connect to the access point. Then, (iv) we reset the packet capture and save the temporary dump to trace the power-on destinations separately. (v) We trigger the device with a set of activities, and finally, (vi) the traffic capture is stopped.
The full procedure produces two PCAP files that are later processed to obtain the list of the remote destinations' domain names and IP addresses.\\ The next step is to analyze the destinations on the obtained lists to distinguish the essential from non-essential ones. The following steps are executed for each destination: first, the traffic from/to the destination is blocked; then, the device is powered off and then immediately back on to reset the device DNS tables. As the device is powered on, it is triggered with a set of operations. During the execution of the different functions, we check if they have been properly executed or not: if all the functions are successful even when the traffic towards the destination is blocked, then the destination is added to the list of non-essential destinations, and the script continues to the next destination maintaining the block to that destination. In cases in which at least one of the tested functions does not produce the expected outcome, the destination is marked as essential for the device operations and can not be blocked: the destination is added to the list of essential destinations and is removed from the block-list during the analysis of successive destinations. The procedure is repeated at least 30 times until an 80\% consensus among the iterations is reached.\\
\begin{table}[t]
\caption{List of IoT devices included in our testbed}
\label{tab:iot_devices_iotrim}
\centering
\resizebox{\columnwidth}{!}{%
\begin{tabular}{|c|c|c|c|c|}
\hline
\textbf{Cathegory} & \textbf{Device} & \textbf{Destinations (Req/Non-Req)} & \textbf{Non-Required Traffic}\\
\hline \hline
  Speaker & Amazon Echo dot 3 & 2/48 & 92.6\%\\
  \hline
  Speaker & Amazon Echo dot 4 & 2/46 & 94.2\%\\
  \hline
   Camera & Yi Pro Home Camera & 1/36 & 91.6\%\\
  \hline
   Appliance & Yeelight Bulb 1S & 1/0 & 0\%\\
  \hline
    Appliance & Tp-Link Kasa Bulb LB120 & 1/3 & 2.6\%\\
 \hline
 Appliance & Tp-Link Tapo Plug P110 & 2/2 & 0.9\%\\
 \hline  
 Hub & SwitchBot Hub Mini & 1/0 & 0\%\\
 \hline
 Video & Roku TV Stick & 1/120 & 81.1\%\\
  \hline
%All & 15 & 20 \\ \hline
\end{tabular}
}
\end{table}

The entire procedure is automated using shell scripts that interact with the MonIoTr software for traffic collection and blocking. We use \textit{Tcpdump} for traffic capturing, while we process the PCAP files using the \textit{tshark} command line tool, part of the \textit{Wireshark} programming suite. We extract all the DNS type A requests and save the requested domain name with the corresponding IP address in the response. With this procedure, we obtain a CSV file with the $(Domain\ name, IP\ address)$ pairs for each destination involved in the device communication from both device boot and activity execution steps. To automatically trigger the device and execute the defined set of functions for the different devices, we use the \texttt{adb}\footnote{https://developer.android.com/tools/adb} interface to control an Android smartphone and interact with the different apps. The smartphone is connected to a separate access point to force communication with IoT devices through cloud services. Devices requiring voice commands are triggered with text-to-speech tools.
For traffic blocking, we implement an \textit{IP blocker} to filter out communication with a specific IP address, while \textit{DNS Override} is implemented to block the translation of a network domain name, thus interrupting the communication with the destination. Finally, to check the results of the function execution, we use device-specific probe scripts to ensure the expected outcome is achieved successfully. For android-triggered functions, the probe scripts are based on screenshot comparisons.
For voice-triggered devices, the automated outcome control is implemented by analyzing the produced traffic based on specific Uplink thresholds, analogously to \cite{mandalari2021blocking}.
\subsection{Testbed}
%Once we have defined a methodology to distinguish the nature of the network destinations contacted by IoT devices, we have all the tools required to build a labeled dataset and proceed with the development of the machine-learning-based solution.
To obtain a labeled network dataset, we need to set up an IoT network with different devices, apply the outlined methodology, and collect the network data.
For this purpose, we selected eight consumer smart home devices, including different categories and brands, as reported in Table \ref{tab:iot_devices_iotrim}. We set an Intel NUC mini PC to act as a Wi-Fi access point, installing the required software and scripts for the collection and analysis pipelines. All the devices are connected to the access point using Wi-Fi technology, and they are connected to the power through a smart plug that allows automated power-on/off procedures. To reduce the number of executions, we define a subset of functions to be tested for each device, and we proceed with the steps of the previously presented methodology to collect the list of all destinations and categorize each one as essential or non-essential for the device. The triggering and probe scripts are adjusted for each device to successfully control and check the outcome of the executions.
%The outcome of this procedure is composed of two lists, one used for essential and one for non-essential destinations. Each destination is represented by the domain name and the IP address (extracted from the DNS requests). If more IP addresses are found for a destination, all of them are included.
The procedure outputs the list of essential/non-essential destinations for each device. The next step is to collect the network data to train/evaluate the classification models: we collect the traffic from the eight devices in our testbed for 4 months, producing one PCAP file per device per day. The devices are periodically rebooted and triggered during traffic collection, alternating periods of activity to inactive periods to have a realistic smart home scenario. %Moreover, to force the device to generate DNS requests periodically, the devices have been randomly rebooted occasionally during the traffic collection.
%As already done in the destination analysis phase, the function trigger is done automatically using \textit{adb} scripts and text-to-speech APIs, and each time an action is executed, a log entry is produced for the device to have ground truth on the generated events. The outcome of the function execution is included in the log for completeness, using the same probe checking as in the previously presented methodology. The log contains the following information: device name, timestamp of the first and last moment of activity, type of activity executed, and success state. \\
%The traffic is saved as raw data in PCAP files: the collection framework is set to produce a single file per device per day, and if the maximum size of 128 MB is reached during one day, the daily traffic file is split into more PCAP files for that device.
Table \ref{tab:iot_devices_iotrim} also reports the results of a preliminary analysis on the nature of the network traffic, reporting the number of essential/non-essential destinations contacted by each device.

\section{Learning Module}
\label{sec:ml-iotrim}
\begin{figure}[t]
    \begin{center}
      \includegraphics[width=0.9\columnwidth]{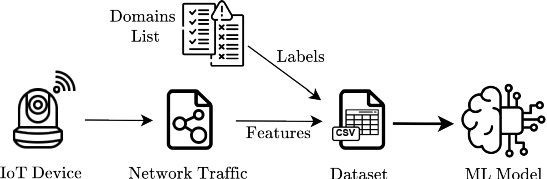}
      \caption{Data processing pipeline from raw traffic to Machine Learning classification.}
      \label{fig:iotrim_model_architecture}
    \end{center}
\end{figure}
The ML-IoTrim learning module consists of the feature-extraction methodology and the subsequent classification. The pipeline is summarized in Figure \ref{fig:iotrim_model_architecture}.
\subsection{Feature Extraction}
For the feature extraction process, we analyze the devices individually. The first step is to extract the DNS packets from all the traffic to obtain the pairs of domain names and the corresponding IP addresses of all the contacted destinations. %This will allow us to model the traffic in terms of destinations rather than IP addresses in case the same destination changes IP addresses over time or uses more IP addresses simultaneously. It is worth pointing out that once the machine learning models are trained, the proposed approach can work even under encrypted DNS traffic, modeling the destination directly with the IP address instead of its domain name. Given the recent release of encrypted DNS solutions (i.e., DNS over HTTP), this represents an important aspect of the adopted solution for long-term support.
Then, each PCAP file for each device is processed individually, and the set of features is extracted with the following pipeline: \begin{itemize}
    \item First, we group packets by destination, checking the previously extracted DNS queries and substituting the IP with the domain name, when possible.
    \item Then, we perform a time-window aggregation, grouping packets involving the same destination into non-overlapping time windows of fixed duration. The value for the window length will be discussed in the results. 
    \item Finally, we compute a set of statistical values on top of packet headers for each time window for each destination for each device: we extract different statistics from packet size, interarrival times, and other TCP/IP information.
\end{itemize} 
To compute the set of features, the packets in each time window are divided into different sets distinguished by protocol (TCP, TLS, UDP, Any) and direction (Uplink, Downlink, Any). For each of the 12 sets obtained with the different combinations, the following features are computed (16 total):
\begin{itemize}
    \item Number of packets
    \item 8 Statistics on packet size: sum, mean, median, standard deviation, minimum and maximum values, first and third quartiles
    \item 7 Statistics on Interarrival Time (IAT): mean, median, standard deviation, minimum and maximum values, first and third quartiles
\end{itemize}
In addition to the 192 features already listed, the following features are computed for each time window:
\begin{itemize}
    \item TCP over UDP packet/byte ratios (2 features)
    \item TLS over TCP packet/byte ratios (2 features)
    \item Uplink over Downlink packet/byte ratios (2 features)
    \item Number of unique UDP and TCP local and remote ports (4 features)
    \item Number of unique TCP and UDP flows, identified with the tuple \textit{$<$device, local port, remote IP, remote port, protocol$>$} (2 features)
\end{itemize}
The obtained set of 204 features is completed with a last column representing the ground truth label (essential/non-essential), assigned using the destinations lists obtained in the first steps of the presented methodology. %In this case, the feature extraction scripts are programmed in Python and produce one CSV file per device per day, each containing the features and the ground-truth label for each time window for each destination for each device.
We compute the different features aggregating packets with time windows of 10, 60, 600, and 3600 seconds to study the impact of such parameter on the classification performance.

\subsection{Machine Learning Classification}
The labeled dataset obtained after feature extraction can be used to train supervised machine-learning models for binary classification. As a first operation, the features are normalized per device to allow an easier generalization of the features between different types of devices and improve the model accuracy.
We refer to different ML models based on decision trees and neural networks. However, we report the results for the two models that obtained the best results: the Random Forest Classifier (RFC) and a custom neural network properly designed on top of the selected features. Given the nature of the aforementioned features, RFC achieves the best results. However, we also investigate the performance of Artificial Neural Networks (ANNs) for the given task to enrich and confirm the outcome of the proposed methodology. We propose a fully-connected neural network with a first linear hidden layer that transforms the input from 204 (the number of available features) values to 128 neurons; then two additional linear hidden layers are used with 64 and 32 neurons. Finally, the last layer has a single neuron, which produces the network's final output, assuming value greater than 0.5 if the destination is essential. %The sigmoid activation function is applied to the output of the last layer, squashing the value between 0 and 1, which is typical for binary classification problems. The decision is finally taken on this value: the destination is detected as essential if greater or equal to 0.5, while it is considered as non-essential otherwise. 
%ReLU is used as an activation function in the three hidden layers, and the Binary Cross Entropy Loss (BCELoss) is used as a loss function. During the training phase, the Adam optimization function is used to update the neural network weights.

\section{Results}
\label{sec:results}

In the first analysis case, we train a global model using the RFC and NN models with the dataset containing information from all eight devices considered in our testbed. The data is split following a time-oriented division to simulate a real-life scenario in which the models are trained and evaluated in different periods. We opt for a train-test-validation split with 0.7, 0.15, and 0.15 ratios. %We analyze and report the results using different values for the time window used to aggregate packets to compute the statistical features. 
\begin{table}[t]
\caption{Global model results under different aggregation windows}
\label{tab:rfc_vs_ann}
\centering
\resizebox{0.9\columnwidth}{!}{%
\begin{tabular}{|c|c|c|c|c|}
\hline
\textbf{Model} & $\bm{w}=\bm{10s}$ & $\bm{w}=\bm{60s}$ & $\bm{w}=\bm{600s}$ & $\bm{w}=\bm{3600s}$\\
\hline
  RFC & 99.88\% & 100\% & 99.89\% & 99.62\%\\
  \hline
  ANN & 98.81\% & 99.46\% & 99.27\% & 98.03\%\\
\hline
\end{tabular}
}
\end{table}

\begin{figure}[t]
    \begin{center}
      \includegraphics[width=\columnwidth]{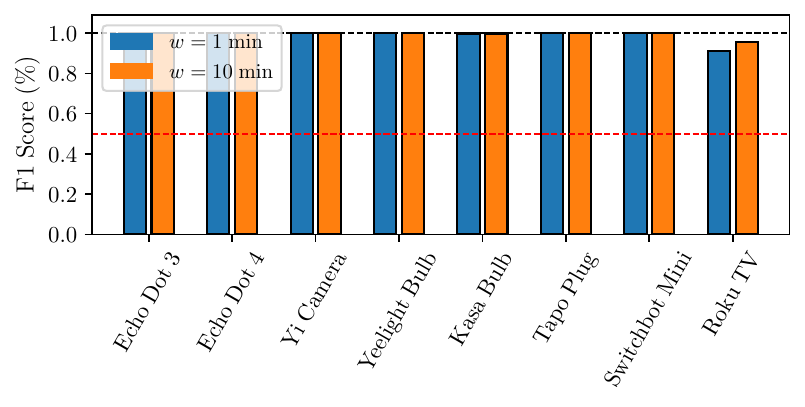}
      \caption{F1 Score of the ANN classification evaluated on single devices}
      \label{fig:global_on_single}
    \end{center}
\end{figure}
We report the F1 Score obtained in the test set by the two models when using different aggregation windows in Table \ref{tab:rfc_vs_ann}. Results show good performance for the two models, with almost optimal classification for the RFC with an F1 score close to 100\%. %for the two models under different aggregation windows Figure \ref{fig:rfc_vs_nn} reports the classification F1 Score and shows that the results are optimal for both algorithms, with a slightly better performance for the Random Forest Classifier.

\begin{figure}[t]
    \begin{center}
      \includegraphics[width=\columnwidth]{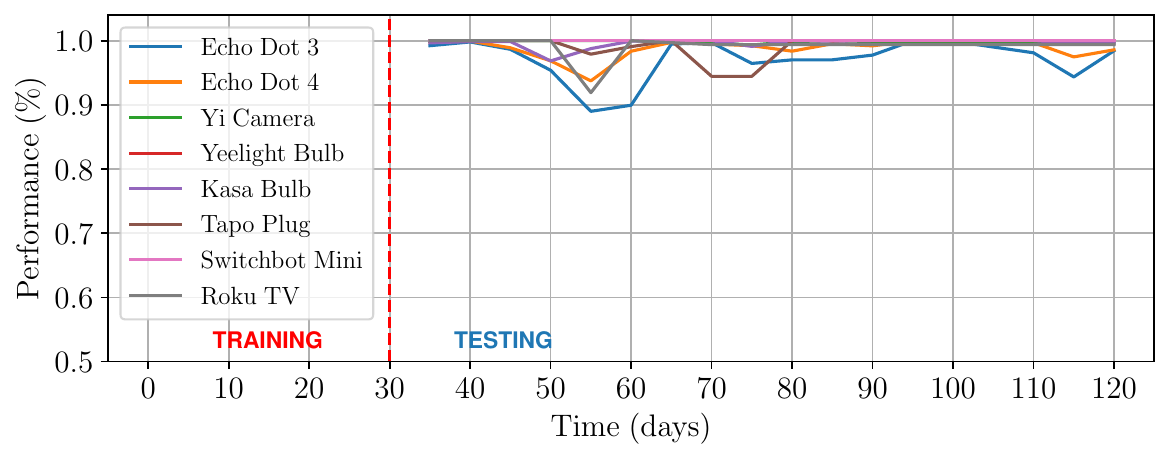}
      \caption{Long-term consistency of the global model over time. The model is trained on the first 30 days of data and evaluated for the remaining 3 months}
      \label{fig:temporal_res}
    \end{center}
\end{figure}

\begin{figure}[t]
    \begin{center}
      \includegraphics[width=\columnwidth]{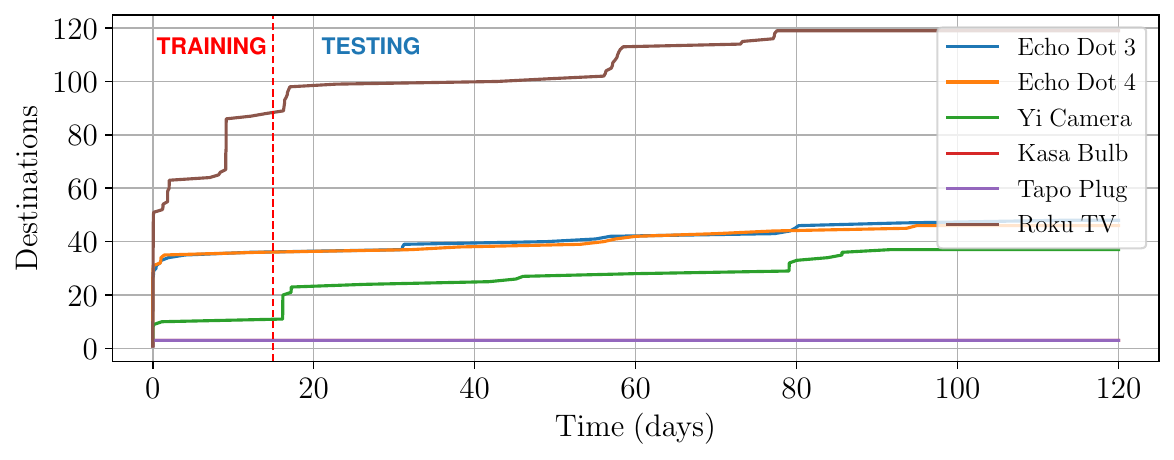}
      \caption{Distribution of unique non-essential destinations contacted over time by the different IoT devices.}
      \label{fig:dest_distribution}
    \end{center}
\end{figure}

\begin{figure}[t]
    \begin{center}
      \includegraphics[width=\columnwidth]{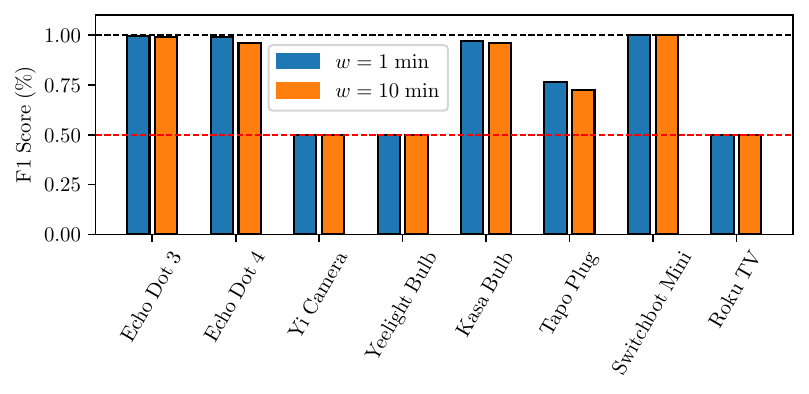}
      \caption{All vs. One RFC classification F1-Score}
      \label{fig:all_vs_one}
    \end{center}
\end{figure}

The value of the aggregation window $w$ slightly influences the classification performance, obtaining better results for $w=60s$ and $w=600s$. 
To further understand the performance of each device, we proceed with the model evaluation when the devices in the test sets are considered individually. Figure \ref{fig:global_on_single} reports the results for the ANN classifier using the two window values of $w=60s$ and $w=600s$. We can observe that for all the devices, the classification is close to the optimum, except for the Roku TV stick. However, the results for this last device are still good, with an F1 Score over 90\%.
Overall, we can state that one single classifier trained on all the devices can effectively distinguish the destinations as essential/non-essential, even if different categories are involved. %However, the training set should contain part of the traffic from each device to classify.
%consistency over time
%As the model does not properly generalize to a correct classification for new devices, 
To validate the proposed model, we proceed with verifying its consistency over time. To do so, we repeat our analysis, this time reducing the training set's size. We train the model using the first 30 days of data from all eight devices, representing 25\% of the full dataset. Then, the model is evaluated with the remaining data, 5 days at a time for each device. We report the F1-Score over each 5-day set, which maintains over 90\% for all the devices, as shown in Figure \ref{fig:temporal_res}. This clearly confirms the model's capacity to handle new data over time.
%new destinations
Finally, we conclude investigating on the model's ability to classify new destinations for the devices. Figure \ref{fig:dest_distribution} reports the number of unique destinations contacted by the devices during the collection period. We excude the Switchbot hub mini and the Yeelight bulb as they are characterized by only essential traffic.
Similarly, Tapo Plug and Kasa Bulb contacted all the destinations at boot time, and are thus excluded from the analysis. The plot highlights that most of the destinations are contacted within the first days of traffic collection. After inspecting the distribution, we selected the first 15 days of traffic for the training set, while we used the remaining part for the evaluation. Destinations present in the training set have been excluded from the evaluation set. The results are optimal for all four devices, as new destinations are properly recognized as non-essential with 100\% accuracy.\\
The presented results have shown optimal classification using a single global model. However, the training set should contain part of the traffic from each device to classify. As a further analysis, we want to discuss how the model adapts to new devices not involved in the training set. For this purpose, we perform an all-vs-one approach and repeat eight iterations in which the models are trained with seven devices and tested with the remaining one. We report the results of this analysis in Figure \ref{fig:all_vs_one}, using the RFC model with an aggregation window of $w=60s$. We observe that the model does not generalize well when tested on devices not considered in the training set. The classification is optimal for some devices, while it is extremely low for others. We see that for devices of similar logic or similar brands, the classification works perfectly: for example, Amazon Echo Dot 3 and 4 obtain an almost perfect classification even if the destinations have different IP addresses and domain names. However, the classification is comparable to random guessing for devices not having a corresponding device with a similar type or same brand in the training set (i.e., the Yi Camera). To understand more about this, we need to extend our study to more devices.


\section{System Implementation}
\label{sec:implementation}
\begin{figure}[t]
    \begin{center}
      \includegraphics[width=\columnwidth]{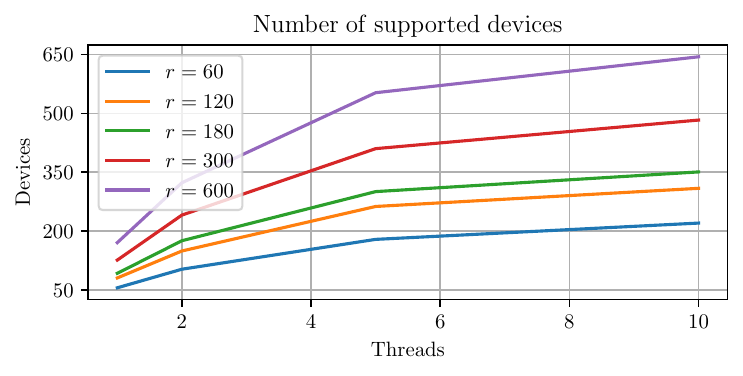}
      \caption{Number of supported devices when using different number of threads to process the data. Values are reported using different rotation periods $r$}
      \label{fig:scalability}
    \end{center}
\end{figure}

We implement the proposed framework in an Intel NUC mini PC acting as an access point for an IoT network. We install the MonIoTr framework to collect and control the traffic, the feature extraction and model inference scripts, and we load the pre-trained models. The capturing scripts are set to save the data in a unique PCAP file for each device and rotate the files after a defined rotation period $r$. After a cycle, the PCAP files for all the devices are processed to extract the aggregated features, and the machine learning models are evaluated on all the time windows for all the destinations found. %The $r$ value needs to be greater or equal to the aggregation window $w$, for which the features get computed, to contain at least one full data window to compute the feature and evaluate the model. 
In cases where the rotation period is greater than the aggregation window, each file may contain more aggregated windows per destination: in such cases, majority voting is performed on the evaluations to take the decision on the destination. After the inference phase, the scripts pass the decision to the IoTrim blocker module: essential destinations cause no special actions in the access point, while if destination is detected as non-essential, it will cause the access point to block it for the specified device, using IP block and DNS override scripts.
%The rotation period can be customized in the traffic collection module, while the data aggregation window value can be specified in the configuration file of the feature-extraction module.
\begin{table}[t]
\caption{Average feature extraction and inference time under different rotation period values}
\label{tab:times_results}
\centering
\begin{tabular}{|c|c|c||c|}
\hline
\textbf{Rotation} & \textbf{Feature} & \textbf{Model} & \textbf{Total Process}\\
\textbf{Period} & \textbf{Extraction} & \textbf{Inference} & \textbf{(50 devices)}\\

\hline \hline
  $r=60s$ & 1.06s & 6.61ms & \textbf{53.6s} \\
  \hline
  $r=120s$ & 1.47s & 6.45ms & \textbf{74.1s} \\
  \hline
  $r=180s$ & 1.93s & 7.27ms & \textbf{96.9s} \\
  \hline
  $r=300s$ & 2.36s & 7.12ms & \textbf{118.5s} \\
  \hline
  $r=600s$ & 3.51s & 7.51ms & \textbf{175.4s} \\
  \hline
  
  %All & 15 & 20 \\ \hline
\end{tabular}
\end{table}
To analyze the scalability and real-time functioning of the system, we evaluate our framework using different rotation periods. To make decisions in real-time, the system should execute the full process in an overall time lower than the rotation period for all the considered devices, to produce the outcome before new data are ready to be processed. We perform several executions using an aggregation window $w=60s$ and different values for the rotation period $r$, and we compute the average feature extraction and inference times during each execution. The system is evaluated with the traffic of 50 active smart cameras. Table \ref{tab:times_results} reports the per-device average feature extraction and inference time, and the cumulative processing time for all the devices. The required time to evaluate the models starting from the PCAP files remains lower than the rotation period in all the considered use cases, showing the real-time capability of the system. We can increase the number of supported devices by executing the process in a multi-threaded fashion. To complete our discussion, we estimate the number of supported devices when multiple threads are used to process the traffic, running the solution with different rotation periods. We process the data from $n$ devices in a multi-threaded architecture with $n$ concurrent threads. We can repeat the procedure $k$ times until the cumulative processing time equals the rotation period. The number of supported devices $N$ is then computed as $N=n\times k$, and is reported in Figure \ref{fig:scalability}, under different values of $r$ and $n$.
Commenting the obtained results, we can conclude that the system can effectively support real-time detection and blocking of non essential traffic scaling to hundreds of devices, if run in a multi-threaded system.

\section{Conclusions}
\label{sec:concl}
The great diffusion of IoT devices around the user is introducing potential security/privacy threats in home environments. This work presented a framework to reduce information exposure in the communication of smart home devices intelligently detecting and blocking non-essential traffic. After presenting the methodology for distinguishing the essential/non-essential destinations to build a labeled dataset containing traces from 8 different devices, we refer to Machine Learning supervised classification techniques to distinguish the destinations. We show that it is possible to detect the non-essential traffic from different categories of devices, only relying on their network traces.
Our solution is time-consistent and can be trained globally to work on single devices. Moreover, the solution can successfully classify new destinations even if not present in the training set. The work concludes with a practical implementation of the intelligent detection framework in a smart home system, in which the access point can dynamically detect and block non-essential destinations from hundreds of devices in a real time fashion.
However, as a negative aspect, we show that the model does not adapt well to devices not present in the training set, which requires further studies in the field. In future research, we plan to extend our testbed with more devices and build per-category models to further investigate the scalability/universality of our solution. To advance the research, we release the code and data publicly.

%\amm{add a comment about scalability of our solution saying more devices can be tested and this is why we publish the code: https://github.com/SafeNetIoT/ML-IoTrim}

\section*{Acknowledgments}
This work was supported by the European Union and the Italian Ministry for University and Research (MUR) through the PRIN project ``COMPACT'' (Mission 4, Component 1, CUP D53D23001340006).

\bibliographystyle{IEEEtran}
\bibliography{bibl}
\end{document}